\documentclass[a4paper]{jpconf}
\usepackage{graphicx}
\begin{document}
\title{Magnetoelectricity in the system $RAl_3(BO_3)_4$ ($R$ = Tb, Ho, Er, Tm)}

\author{K.-C. Liang$^1$, R. P. Chaudhury$^2$, B. Lorenz$^1$, Y. Y. Sun$^1$, L. N. Bezmaternykh$^3$, I. A. Gudim$^3$, V. L. Temerov$^3$ and C. W. Chu$^{1,4}$}

\address{$^1$ TCSUH and Department of Physics, University of Houston, Houston, TX 77204, USA}
\address{$^2$ Intel Corporation, 2501 N.W. 229th St., Hillsboro, OR 97124, USA}
\address{$^3$ Institute of Physics, Siberian Division, Russian Academy of Sciences, Krasnoyarsk, 660036, Russia}
\address{$^4$ Lawrence Berkeley National Laboratory, 1 Cyclotron Road, Berkeley, CA 94720, USA}

\ead{blorenz@uh.edu}

\begin{abstract}
The magnetoelectric effect in the system $RAl_3(BO_3)_4$  ($R$ = Tb, Ho, Er, Tm) is investigated between 3 K and room temperature and at magnetic fields up to 70 kOe. We show a systematic increase of the magnetoelectric effect with decreasing magnetic anisotropy of the rare earth moment. A giant magnetoelectric polarization is found in the magnetically (nearly) isotropic $HoAl_3(BO_3)_4$. The polarization value in transverse field geometry at 70 kOe reaches 3600 $\mu C/m^2$ which is significantly higher than reported values for the field-induced polarization of linear magnetoelectric or even multiferroic compounds. The results indicate a very strong coupling of the f-moments to the lattice. They further indicate the importance of the field-induced ionic displacements in the unit cell resulting in a polar distortion and a change in symmetry on a microscopic scale. The system $RAl_3(BO_3)_4$ could be interesting for the technological utilization of the high-field magnetoelectric effect.
\end{abstract}

\section{Introduction}
The magnetoelectric (ME) effect in materials has been a matter of investigation for more than a century. The original conjecture of Pierre Curie \cite{curie:1894} about the existence of materials that can be electrically polarized by an external magnetic field was later considered by Peter Debye \cite{debye:26} who defined the term "magnetoelectric". It took another few decades, shortly after Dzyaloshinskii \cite{dzyaloshinskii:59} theoretically predicted the ME effect in antiferromagnetic $Cr_2O_3$, that the electric-field induced magnetization \cite{astrov:60} and the magnetic-field induced polarization \cite{folen:61} had been found at low temperatures in $Cr_2O_3$. This experimental demonstration of the ME effect in materials initiated several decades of extensive studies and the search for other compounds with larger ME coupling \cite{schmid:03}.
The current record holder of the linear magnetoelectric effect ($P\propto H$, $P$ electrical polarization, $H$ magnetic field) is $TbPO_4$ with a coefficient of 730 ps/m at 1.5 K and fields below 8 kOe \cite{rivera:09}.

The ME coupling can be derived from thermodynamic relations by expanding the free energy in terms of the magnetic and electric fields \cite{schmid:03}. The lowest-order term that involves magnetic as well as electric fields, $\alpha_{ij}E_iH_j$, defines the linear magnetoelectric tensor, $\alpha_{ij}$. The magnitude of $\alpha_{ij}$ is usually small, for $Cr_2O_3$ it is not larger than 5 ps/m and for the recently discovered high-temperature magnetoelectric hexaferrite the maximum of $\alpha_{ij}$ is 250 ps/m \cite{kitagawa:10}. More importantly, at higher magnetic fields, the linear magnetoelectric effect breaks down (above 8 kOe in $TbPO_4$) \cite{kahle:86} or decreases rapidly (above 5 kOe in hexaferrites). This limits the magnitude of the electrical polarization that can be induced by magnetic fields to relatively low values. Therefore, the bilinear magnetoelectric coupling of the form $\beta_{ijk}E_iH_jH_k$ in the free energy expansion may become more important and even dominate the ME effect at higher magnetic fields.

The search for materials with larger ME interactions has been extended recently to the study of multiferroic materials \cite{fiebig:05,eerenstein:06,tokura:07}. In multiferroics, the electrical polarization is induced by certain types of magnetic orders as a secondary order parameter resulting in improper ferroelectricity. Sizable values of the polarization have indeed been achieved in the ferroelectric state (e.g. 2500 $\mu C/m^2$ in $DyMnO_3$) \cite{kimura:05}, however, the tunability in magnetic fields is rather limited and frequently associated with phase transitions in the multiferroic state within a narrow field or temperature range. Therefore, attention has shifted to alternative classes of materials with large ME polarizations such as the rare earth iron borates, $RFe_3(BO_3)_4$ \cite{zvezdin:05,yen:06,krotov:06,zvezdin:06b}.

The $RFe_3(BO_3)_4$ compounds crystallize in the trigonal huntite structure, space group $R$32 (No. 155) \cite{joubert:68}, and experience a structural transition at $T_S$ to the space group $P$3$_1$21 (No. 152) with sizable anomalies in the specific heat \cite{hinatsu:03} and the dielectric constant \cite{yen:06}. The structural transition temperature decreases from 430 K ($R$=Ho, Y) to about 90 K ($R$=Eu) with increasing rare earth ionic radius \cite{hinatsu:03}. At lower temperatures, the magnetic exchange interactions between the iron spins results in an antiferromagnetic order of the Fe moments at $T_N<$40 K. The iron spins are relatively close along the $c$-axis since the $FeO_6$ octahedra form edge-sharing helical chains along $c$ (Fig. 1). Direct as well as super exchange interactions are supposed to contribute to the coupling of the spins. Different chains are linked through $BO_3$ units (as labeled in Fig. 1) and magnetic exchange between two Fe moments of neighboring chains involves two oxygen ions of the $BO_3$ triangle. The magnetic rare earth ions are embedded between the $FeO_6$ chains and their f-moment couples strongly to the d-spin of the iron at lower temperatures. The existence and coupling of two magnetic species results in complex phase diagrams with spin rotation transitions of the Fe moments induced by the strong magnetic anisotropy of the rare earth ions \cite{yen:06,chaudhury:09}.
\begin{figure}
\begin{center}
%\begin{minipage}{16pc}
\includegraphics[angle=0, width=25pc]{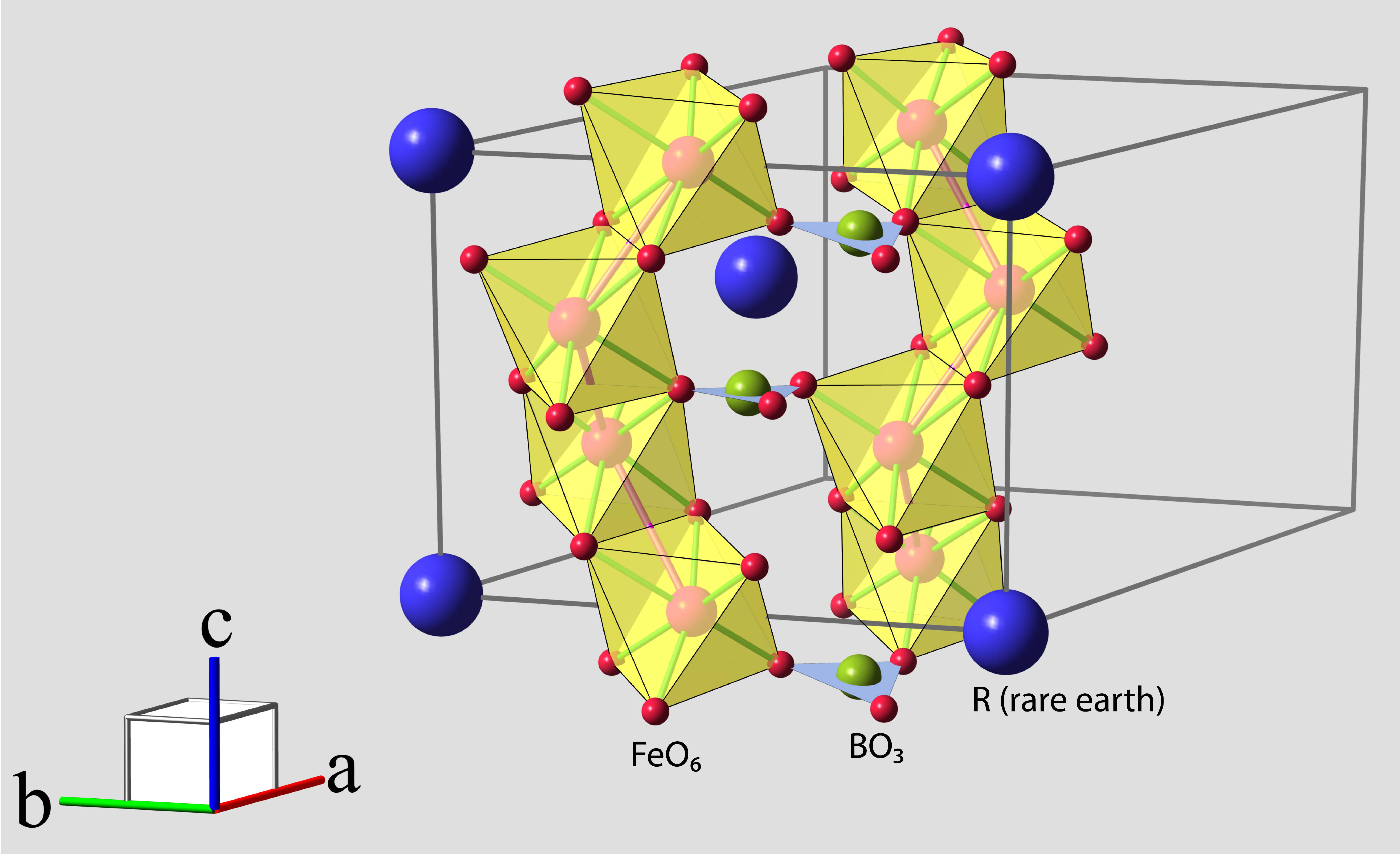}
\caption{\label{label}Part of the structure of $RFe_3(BO_3)_4$ highlighting two helical chains of edge sharing $FeO_6$ polyhedra.}
%\end{minipage}\hspace{4pc}%
\end{center}
\end{figure}
The magnetoelectric properties of the $RFe_3(BO_3)_4$ system have been investigated in detail \cite{kadomtseva:10} and the largest ME effect was reported for $NdFe_3(BO_3)_4$ at high fields up to 20 Tesla \cite{zvezdin:06b}. However, it is not clear whether the large ME effect is associated with the rare earth moment of the iron spin and their coupling to the lattice. The wealth of magnetic interactions between d-electron spins and rare earth moments results in some instances in very complex ME properties as recently shown in $HoFe_3(BO_3)_4$ and the mixed crystal system $Ho_{1-x}Nd_xFe_3(BO_3)_4$ \cite{chaudhury:09,chaudhury:10c}. This makes a physical understanding of the magnetoelectricity and magnetic properties of this system very difficult.

We have therefore synthesized and studied the isostructural compound system $RAl_3(BO_3)_4$ where the transition  metal (Fe) was completely replaced by the nonmagnetic aluminum. Substituting Al for Fe substantially reduces the magnetic correlations since the remaining f-moments are more than 6 ${\AA}$ apart (see the partial structure shown in Fig. 1). This contribution addresses the question of whether or not the simplified magnetic system still gives rise to a sizable ME effect and how it may depend on the magnetic anisotropy of different rare earth ions.

\section{Synthesis and experimental methods}
Single crystals of the system $RAl_3(BO_3)_4$ ($R$=Tb, Ho, Er, Tm) were grown from the seed crystals as described earlier \cite{temerov:08}. The samples of sizes from 5 to 8 mm display well developed growth faces reflecting the trigonal symmetry. In the following we will use an orthogonal coordinate system with $x$ and $z$ parallel to the hexagonal $a$- and $c$-axes, respectively, and $y$ perpendicular to $x$ and $z$. A selection of crystals is shown in Fig. 2. The crystals were cut and shaped according to the demands of the magnetic and magnetoelectric studies. The magnetization was measured in a Superconducting Quantum Interference Device (SQUID) magnetometer (MPMS, Quantum design) with an applied field of 1000 Oe. The magnetoelectric measurements were conducted in a Physical Property Measurement System (PPMS, Quantum Design) for magnetic field and temperature control. The field-induced polarization was determined by integrating the magnetoelectric current arising from a change of polarization while sweeping the magnetic field at a constant speed of 200 Oe/s. The polarization was measured along different crystallographic orientations with longitudinal as well as transverse field directions.

\begin{figure}
\begin{center}
%\begin{minipage}{16pc}
\includegraphics[angle=0, width=30pc]{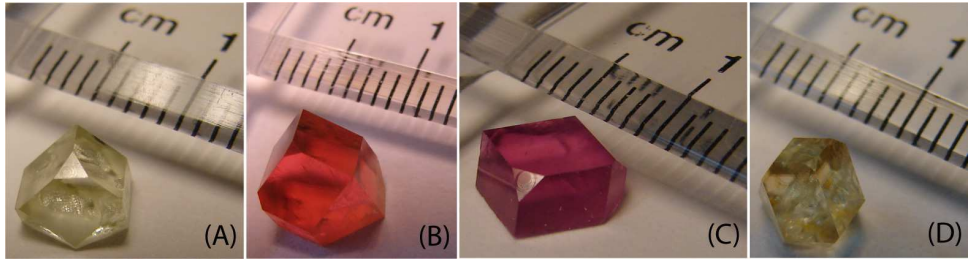}
\caption{\label{label}Single crystals of the compound family $RAl_3(BO_3)_4$. (A) $R$=Tb, (B) $R$=Ho, (C) $R$=Er, and (D) $R$=Tm}
%\end{minipage}\hspace{4pc}%
\end{center}
\end{figure}

\section{Results and discussion}
\subsection{Magnetic properties}
Fig. 3 shows the magnetization for the four compounds along $x$ and $z$ directions. The anisotropy of the magnetism is determined by the properties of the rare earth ion. $TbAl_3(BO_3)_4$ exhibits the strongest uniaxial anisotropy with $\chi_z$ exceeding the in-plane $\chi_x$ by more than a factor of 250 at low temperatures (Fig. 3a). In contrast, $ErAl_3(BO_3)_4$ shows the largest easy plane anisotropy, followed by $TmAl_3(BO_3)_4$ (Figs. 3b and 3c, respectively). $HoAl_3(BO_3)_4$ is magnetically nearly isotropic with a minute preference of the $c$-axis (Fig. 3d). The anisotropy ratio $\chi_z/\chi_x$ is 1.2 over a large temperature range and increases slightly to about 3 at the lowest temperatures. These results are in contrast to an earlier report \cite{neogy:96} which found $\chi_x$ slightly larger than $\chi_z$ and predicted a reversal of $\chi_z/\chi_x$ at 2 K. For a quantitative comparison, the values of the anisotropy ratios at 5 K are given in Table 1 for all compounds. The effective magnetic moments, as determined from a Weiss fit of the high-temperature inverse susceptibilities, are also included in Table 1.

\begin{figure}
\begin{center}
%\begin{minipage}{16pc}
\includegraphics[angle=0, width=36pc]{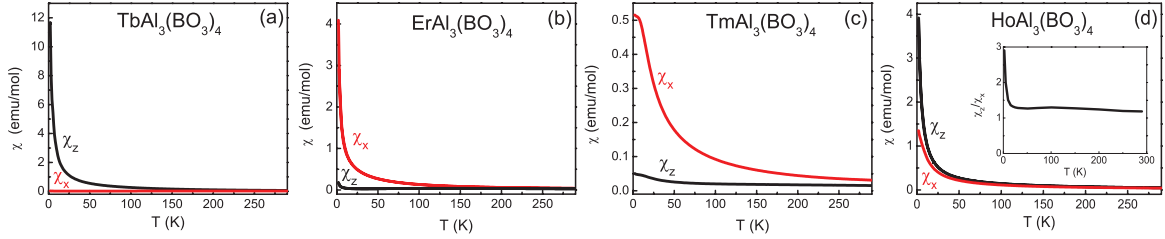}
\caption{\label{label}In-plane ($\chi_x$) and $c$-axis ($\chi_z$) magnetic susceptibilities of $RAl_3(BO_3)_4$.
The inset to (d) reveals the nearly isotropic magnetic properties of $HoAl_3(BO_3)_4$}
%\end{minipage}\hspace{4pc}%
\end{center}
\end{figure}

\begin{center}
\begin{table}[h]
\caption{Magnetic anisotropy at 5 K and effective magnetic moment of $RAl_3(BO_3)_4$.}
%\footnotesize\rm
\centering
\begin{tabular}{@{}*{7}{l}}
\br
Rare earth ion&Tb&Er&Tm&Ho\\
\mr
$\chi_z/\chi_x$&250&0.045&0.1&1.9\\
$\mu_{eff}$ ($\mu_B$)&9.7&9.3&7.4&10.6\\
%\verb"letterpaper"&Set the paper size and margins for US letter paper.\\
\br
\end{tabular}
\end{table}
\end{center}

\subsection{Magnetoelectric properties}
The magnetoelectric properties of the system $RAl_3(BO_3)_4$ depend on the splitting of the f-orbitals in the crystal field, the occupation of the f-levels, and the coupling to the lattice. Since the noncentrosymmetric space group R32 is a non-polar group, the external magnetic field has to introduce a polar distortion to give rise to a macroscopic electrical polarization. This polarization can be significant, as shown earlier for $R$=Tm, Ho \cite{chaudhury:10b,liang:11}. The role of the magnetic anisotropy, which is also determined by the occupied crystal field levels, and its relation to the magnetoelectric couplings is of interest and is studied in detail for Tb, Er, Tm, and Ho in the $RAl_3(BO_3)_4$ structure.

Measurements of the magnetoelectric current of $TbAl_3(BO_3)_4$ along the in-plane $x$- and out-of-plane $z$-directions in longitudinal and transverse magnetic fields have not yielded any magnetoelectric signal within the resolution of the measurement. Therefore, we conclude that $TbAl_3(BO_3)_4$ with the largest uniaxial magnetic anisotropy is either not magnetoelectric or the ME effect is very small. A similar reduction of the ME effect due to strong uniaxial anisotropy of the rare earth moment was discussed in the isostructural $RFe_3(BO_3)_4$ system and it was attributed to the small in-plane components of the f-moments for $R$=Pr, Tb, Dy \cite{kadomtseva:10}. Although the $RFe_3(BO_3)_4$ system is more complex due to the presence of Fe-spins, their magnetic order, and their exchange coupling with the rare earth, it appears conceivable that the phenomenological model derived for the rare earth iron borates could qualitatively explain the missing ME effect of $TbAl_3(BO_3)_4$.

On the contrary, the nearly isotropic $HoAl_3(BO_3)_4$ and the easy-plane magnets $ErAl_3(BO_3)_4$ and $TmAl_3(BO_3)_4$ have been found to display a sizable magnetoelectric polarization, with the orientation and magnitude depending on the rare earth ion. For the three ME compounds we found the largest ME effect in transverse field orientations with the electrical polarization measured along the $x$-axis and the magnetic field applied in $y$-direction. This transverse ME polarization is shown at different temperatures in Fig. 4. The sign of the polarization is referenced to the longitudinal polarization $P_x(H_x)$ which was chosen to be positive. The transverse $P_x(H_y)$ has the opposite sign in all cases (Fig. 4). For $ErAl_3(BO_3)_4$, the magnitude of $P_x(H_y)$ increases to 140 $\mu C/m^2$ at the lowest temperature of 3 K and at a field of 70 kOe. At higher temperatures, the polarization values are diminished, as shown in Fig. 4a. However, the magnetoelectric effect is still significant at temperatures as high as 100 K. This distinguishes the $RAl_3(BO_3)_4$ system from the rare earth iron borates where a sizable ME effect was only detected below the Ne\'{e}l temperature ($\sim$ 40 K) of the Fe spins \cite{kadomtseva:10}.

The magnitude of the transverse polarization increases even further in $TmAl_3(BO_3)_4$ (up to 750 $\mu C/m^2$, Fig. 4b) and $HoAl_3(BO_3)_4$ (up to 3600 $\mu C/m^2$, Fig. 4c) at 3 K and 70 kOe. These values exceed the ME polarizations obtained at the same magnetic fields in the $RFe_3(BO_3)_4$ systems significantly. The maximum of 3600 $\mu C/m^2$ in $HoAl_3(BO_3)_4$ appears to top the reported field-induced polarizations in the known magnetoelectric compounds and even in multiferroics, e.g. $DyMnO_3$ ($P_{max}$=2500 $\mu C/m^2$) \cite{kimura:05}. This result leads us to conclude that the d-electron spin of the Fe ion in $RFe_3(BO_3)_4$ and the antiferromagnetic order does not facilitate the ME effect. It rather seems to be detrimental, preventing large field-induced polarization values. It is also interesting that the ME polarization in the system $RAl_3(BO_3)_4$ is correlated with the magnetic anisotropy of the rare earth ion. $P_x(H_y)$ increases when the anisotropy decreases. This behavior indicates that the components of the f-moment in the hexagonal plane as well as perpendicular to the plane are both essential for the ME properties. The magnetoelectric effect decreases whenever one component becomes small and the anisotropy (uniaxial or in-plane) increases. Therefore, the magnetically isotropic rare earth ion gives rise to the largest polarization values.

The ME effect is associated with a field-induced structural distortion to a polar structure since the zero-field space group $R32$ is non-polar. This macroscopic distortion was detected in the $RAl_3(BO_3)_4$ system through magnetostriction measurements and was found significant \cite{chaudhury:10b,liang:11}. Unfortunately, macroscopic magnetostriction data cannot reveal the microscopic displacements of the ions giving rise to the ME polarization. The magnetic field effects on the microscopic structure and the nature of the distortions could be studied through scattering (X-ray, neutron) methods in magnetic fields. Those experiments will eventually lead to a more comprehensive understanding of the ME effects in the $RAl_3(BO_3)_4$ class of materials.

\begin{figure}
\begin{center}
%\begin{minipage}{16pc}
\includegraphics[angle=0, width=36pc]{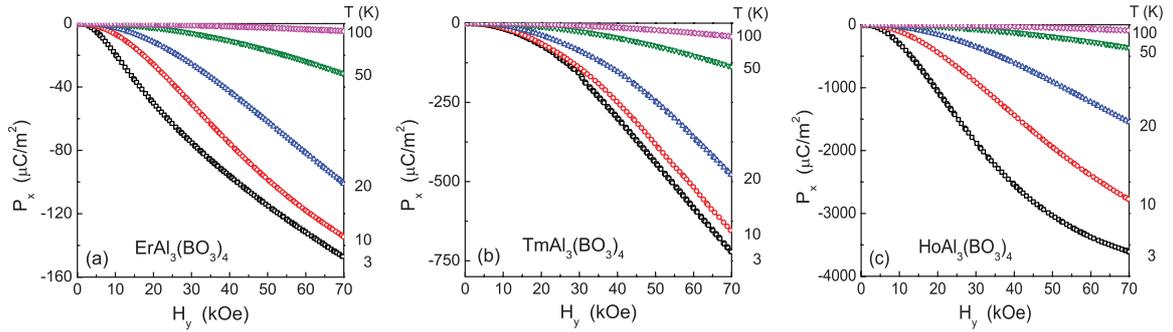}
\caption{\label{label}Transverse magnetoelectric polarizations $P_x(H_y)$ of $RAl_3(BO_3)_4$.}
%\end{minipage}\hspace{4pc}%
\end{center}
\end{figure}

\section{Summary}
We have studied the magnetoelectric effect in the system $RAl_3(BO_3)_4$ for rare earth ions Tb, Er, Tm, and Ho with different magnetic anisotropies. While no magnetoelectric effect was found in the easy axis magnet $TbAl_3(BO_3)_4$, the easy plane and nearly isotropic magnets $ErAl_3(BO_3)_4$, $TmAl_3(BO_3)_4$, and $HoAl_3(BO_3)_4$ exhibit a large magnetoelectric polarization in external fields, exceeding values reported for the sister compound system $RFe_3(BO_3)_4$ significantly in the same magnetic field range. We show that the transverse magnetoelectric effect is largest for all compounds and it increases with the decrease of the magnetic anisotropy. X-ray or neutron scattering experiments in magnetic fields are proposed to elucidate the microscopic distortions giving rise to the large magnetoelectric effect.

\ack This work is supported in part by the T.L.L. Temple Foundation, the J. J. and R. Moores Endowment, the United States Air Force Office of
Scientific Research, the Department of Energy, and the State of Texas through TCSUH.
\section*{References}

\providecommand{\newblock}{}

%\bibliography{HMO}

%\begin{thebibliography}{9}
%\bibitem{iopartnum} IOP Publishing is to grateful Mark A Caprio, Center for Theoretical Physics, Yale University, for permission to include the {\tt iopart-num} \BibTeX package (version 2.0, December 21, 2006) with  this documentation. Updates and new releases of {\tt iopart-num} can be found on \verb"www.ctan.org" (CTAN).
%\end{thebibliography}

\end{document}